\documentclass[prd,preprint,nofootinbib,amsmath,amssymb]{revtex4}

\usepackage{graphicx}
\usepackage{color}

\newcommand{\tr}{\mathrm{tr}}
\newcommand{\R}{\mathcal{R}}
\newcommand{\Z}{\mathcal{Z}}
\newcommand{\Dslash}{\not\hspace{-.3em}D}

\begin{document}

\title{Gell-Mann-Oaks-Renner-like relation in IR-conformal gauge theories}

\author{Agostino Patella}
\affiliation{CERN\\CH-1211 Geneva 23, Switzerland}
\email{agostino.patella@cern.ch}

\begin{abstract}
A generalization of the Gell-Mann-Oaks-Renner relation to the case of infrared-conformal gauge theories is discussed. The starting point is the chiral Ward identity connecting the isovector pseudoscalar susceptibility to the chiral condensate, in a mass-deformed theory. A renormalization-group analysis shows that the pseudoscalar susceptibility is not saturated by the lightest state, but a contribution from the continuum part of the spectrum survives in the chiral limit. The computation also shows how infrared-conformal gauge theories behave differently, depending on whether the anomalous dimension of the chiral condensate be smaller or larger than 1. An application to lattice simulations is briefly discussed.

\vspace{2cm}

\noindent
CERN-PH-TH/2011-144
\end{abstract}

\maketitle

\section{Introduction}

The goal of this paper is to study some analytical aspects of gauge theories in the conformal window related to explicit breaking of the chiral symmetry. I have in mind $SU(N)$, $SO(N)$ and $Sp(N)$ gauge theories, whose fundamental degrees of freedom are gluons and minimally coupled Dirac fermions (\textit{flavors}) transforming under the gauge symmetry in a generic representation of the gauge group.

For few flavors, chiral symmetry is expected to be spontaneously broken in such theories by the fermion condensate $\langle \bar{\psi} \psi \rangle$. At large enough number of flavors, asymptotic freedom is destroyed. It is commonly believed that a window (\textit{conformal window}) for the number of flavors exists in between these two regimes in which chiral symmetry is restored and the theory becomes asymptotically scale invariant at large distances. The existence of the conformal window has been conjectured by Banks and Zaks~\cite{Banks:1981nn}, who showed that the two-loop running coupling constant displays an infrared (IR) fixed point if the number of flavors is properly chosen. The computation of Banks and Zaks can be trusted only in the large-$N$ Veneziano limit of an $SU(N)$ theory. If $N_f$ is the number of flavors of fundamental fermions, then the ratio $N_f/N$ is kept fixed, and it is a continuum parameter that can be tuned arbitrarily close to its value at which asymptotic freedom is destroyed. The existence and size of the conformal window have been investigated by means of several analytical methods~\cite{Appelquist:1988yc, Cohen:1988sq, Miransky:1996pd, Ryttov:2007cx, Appelquist:1999hr, Poppitz:2009uq, Armoni:2009jn}, based on educated but often uncontrolled approximations.

Far away from the Bank-Zaks fixed point, these theories become strongly coupled and analytical investigation becomes inherently difficult. However scale invariance is a quite powerful tool. The anomalous Ward identity for the scale transformations allows to derive very general and interesting relationships, also when scale invariance is broken in a controlled fashion by dimensionful parameters in the action (like a mass term or a finite volume). Even though the techniques proposed in this paper are widely known and used in several areas of theoretical physics, their potential has not been fully exploited in the particular case of gauge theories in the conformal window. A recent analysis can be found for instance in~\cite{DelDebbio:2010hx}. (Broken) scale invariance can be also investigated in the Hamiltonian formalism, studying the action of the trace of the stress-energy tensor on physical states. This complementary approach has been recently exploited in~\cite{DelDebbio:2010jy, DelDebbio:2010ze}.

Before moving to the main point of this paper, it is worth reporting that many numerical investigations (via lattice simulations) of the conformal window has been produced in the past few years.\footnote{For a recent review, see~\cite{DelDebbio:2010zz}. A selection of recent results follows: \cite{Appelquist:2009ka, Appelquist:2009ty, Appelquist:2007hu, Appelquist:2011dp, Fodor:2009wk, Fodor:2009ar, Fodor:2011tw, Fodor:2011tu, Hasenfratz:2009ea, Hasenfratz:2010fi, Hasenfratz:2011xn, DeGrand:2009mt, DeGrand:2009hu, DeGrand:2011cu, DeGrand:2011qd, DeGrand:2010na, DeGrand:2008kx, Shamir:2008pb, Svetitsky:2009pz, Deuzeman:2010zz, Deuzeman:2009mh, Deuzeman:2008sc, Hietanen:2009zz, Hietanen:2009az, Hietanen:2008mr, Karavirta:2011mv, Kogut:2010cz, Kogut:2011ty, Bursa:2010xn, Bursa:2009we, DelDebbio:2010hu, DelDebbio:2010hx, DelDebbio:2009fd, DelDebbio:2008zf}.
} The interest of the lattice community has been boosted by the proposal that theories inside or close to the conformal window can be suitable for building walking or conformal technicolor models for electroweak symmetry breaking~\cite{Eichten:2011sh, Appelquist:1986an, Sannino:2004qp, Andersen:2011yj, Luty:2004ye, Evans:2010ed, Sannino:2008nv}. Aside this possible application, the study of gauge theories in the conformal window is an interesting subject on its own, which will eventually lead to a better and deeper understanding of strongly-coupled gauge theories in general. While setting up numerical simulations and interpreting data, one has always to keep in mind that gauge theories in the conformal window are deeply different from QCD and most of the wisdom and intuition we might have from QCD does not work in general. Analytical works are necessary in order to develop new tools for guiding numerical simulations.

\medskip

For few flavors, chiral symmetry breaking is signaled by a nonvanishing value of the chiral condensate $\Sigma=-\langle \bar{\psi}\psi \rangle/N_f$. The Goldstone bosons (pions) appear as a pole in the longitudinal part of the current that generates the chiral symmetry (isovector axial current). The amplitude of the pole is related to the pion electroweak decay constant $F_{PS}$.\footnote{More precisely, if $A_{\mu} = \bar{u} \gamma_5 \gamma_\mu d$ is the isovector axial current and $|\pi,p_\mu\rangle$ is the state of a $\bar{u}d$ pion with momentum $p_\mu$ and relativistic normalization
$$
\langle \pi, p_\mu | \pi, q_\mu \rangle = 2p_0 (2 \pi)^3 \delta^3(\mathbf{p}-\mathbf{q}) \ ,
$$
then the pion decay constant is defined as:
$$
\langle 0 | A_\mu(0) | \pi, p_\mu \rangle = F_{PS} p_\mu \ .
$$
}
If the quarks have a small mass $m$, then chiral symmetry is explicitly broken and the pions acquire a mass $M_{PS}$. The Gell-Mann-Oaks-Renner (GMOR) formula~\cite{GellMann:1968rz,Gasser:1982ap} relates the quark and pion masses, the chiral condensate and the pion decay constant close to the chiral limit:
\begin{gather}
m \Sigma = F_{PS}^2 M_{PS}^2 \ .
\label{eq:GMOR}
\end{gather}
The GMOR relation is an extremely useful tool in lattice simulations for identifying the chiral region, and for defining a chiral condensate which is free of additive renormalization with Wilson fermions~\cite{Giusti:1998wy}.

This work investigates how the GMOR relation generalizes in the conformal window. The starting point is the integrated Ward identity for the chiral symmetry applied to the isovector pseudoscalar (PS) density $P = \bar{u} \gamma_5 d$:
\begin{gather}
\int \langle P^\dag(x) P(0) \rangle d^4x = - \frac{\langle \bar{\psi}\psi \rangle}{mN_f} \ .
\label{eq:ward}
\end{gather}
When chiral symmetry is spontaneously broken, the pion susceptibility is divergent in the chiral limit. The divergence is saturated by the single-pion state propagating in the correlator (\textit{pole dominance}). Since the chiral condensate is finite in the chiral limit, the rhs of the Ward identity is also divergent in the chiral limit. The GMOR relation is derived by equating the divergences in both sides.

In case of IR-conformality, I will show that the pole dominance does not hold. At the leading order in the chiral limit, both the pole and the continuum spectrum contribute to the isovector PS susceptibility. Moreover the isovector PS susceptibility is divergent in the chiral limit only if the anomalous dimension of the chiral condensate is in the range $1<\gamma_*<2$.\footnote{Notice that the anomalous dimension of the chiral condensate is positive, but contrained to be less than 2 by unitarity~\cite{Mack:1975je,Grinstein:2008qk,Rattazzi:2008pe}.} Once properly regularized, the chiral condensate behaves in the chiral limit like:
\begin{gather}
-\langle \bar{\psi}\psi \rangle \simeq A m^{\frac{3-\gamma_*}{1+\gamma_*}} \ ,
\end{gather}
and the rhs of eq.~\eqref{eq:ward} is divergent. In this case a matching of the divergences of the Ward identity~\eqref{eq:ward} is still possible, but the GMOR relation is modified schematically as follows:
\begin{gather}
m \Sigma = F_{PS}^2 M_{PS}^2 + \textrm{continuum-spectrum contribution} \ .
\end{gather}

If the anomalous dimension is less than 1, the isovector PS susceptibility does not diverge in the chiral limit. Moreover the non-analytical term is subleading in the chiral condensate. For theories with a $m \to -m$ symmetry one gets:
\begin{gather}
-\langle \bar{\psi}\psi \rangle  \simeq \alpha_{\bar{\omega}} m + A m^{\frac{3-\gamma_*}{1+\gamma_*}} \ .
\end{gather}
This problem can be circumvented by taking the first and second derivatives of the Ward identity~\eqref{eq:ward} with respect to the mass:
\begin{gather}
\int \langle \bar{\psi}{\psi}(y) P^\dag(x) P(0) \rangle_c \ d^4x \, d^4y = - \frac{\partial}{\partial m} \left[ \frac{\langle \bar{\psi}\psi \rangle}{mN_f} \right] \ ,
\label{eq:ward_first} \\
\int \langle \bar{\psi}{\psi}(z) \bar{\psi}{\psi}(y) P^\dag(x) P(0) \rangle_c \ d^4x \, d^4y \, d^4z = - \frac{\partial^2}{\partial m^2} \left[ \frac{\langle \bar{\psi}\psi \rangle}{mN_f} \right] \ .
\label{eq:ward_second}
\end{gather}
If $1/3<\gamma_*<1$ then the integral of the 3-point function $ \langle \bar{\psi}{\psi}(y) P^\dag(x) P(0) \rangle_c $ diverges in the chiral limit, therefore one can obtain a generalized GMOR relation by matching the divergences of the first derivative of the Ward identity~\eqref{eq:ward_first}. Finally if $0<\gamma_*<1/3$ also the 3-point function is integrable in the chiral limit, and one has to consider the second derivative of the Ward identity~\eqref{eq:ward_second}.

\medskip

It is well known that the Ward identity~\eqref{eq:ward} contains UV-divergences. Section~\ref{sec:notation} is devoted to present the basic notations used in this paper, and to elaborate a formulation of the Ward identity~\eqref{eq:ward} that is free from UV-divergences. The outcome will be the \textit{master equation}, which will be applied to IR-conformal theories deformed with a mass term for the fermions. The basic tool for investigating the implications of IR-conformality is represented by the renormalization group (RG) equations. The scaling properties of the relevant observables (isovector PS spectral density and Dirac eigenvalue density) are analyzed in detail in section~\ref{sec:RG}. This is a technical but crucial step for identifying the IR-divergences in the Ward identity (and its derivatives). The generalized GMOR relation will be obtained for $1<\gamma_*<2$ in section~\ref{sec:GMOR_large}, and for $0<\gamma_*<1$ in section~\ref{sec:GMOR_small}. Finally a possible application to lattice simulations is discussed in section~\ref{sec:lattice}.

\section{Basic notations}
\label{sec:notation}

\subsection{Regulated chiral Ward identity}

As already mentioned in the introduction, the chiral Ward identity~\eqref{eq:ward} suffers of UV divergences. In fact the isovector\footnote{
Since I am interested in this paper only in the \textit{isovector} PS channel, I will omit the word \textit{isovector} from now on.
} PS correlator $C_{PS}(x) = \langle P^\dag(x) P(0) \rangle$ diverges like $1/|x|^6$ (up to logarithms) at short distances, which gives a quadratic divergence when integrated over the spacetime. In the chiral limit, it may suffer also of IR divergences. In a generic power-law scenario in which $C_{PS} \simeq 1/|x|^{\alpha}$ at large distances, the chiral Ward identity would contain an IR divergence for $\alpha \le 4$ (notice that in case of chiral symmetry breaking $\alpha=2$). The goal of this section is to derive a formulation of the chiral Ward identity in which UV and IR divergences are separated. This will be achieved in two steps.

The \textit{first step} consists in expanding both the PS correlator and the chiral condensate in eigenmodes of the Dirac operator. Let $X_a(x)$ be the eigenvector of the Dirac operator with eigenvalue $\omega_a$ (at fixed gauge background and finite volume) and let $\rho(\omega,m)$ be the eigenvalue density defined as:
\begin{equation}
\rho(\omega,m) = \lim_{L \to \infty} \frac{1}{L^4} \sum_a \langle \delta(\omega-\omega_a) \rangle \ .
\label{eq:rho_def}
\end{equation}
The eigenvalue density $\rho$, as all the other observables, is a function of the renormalized fermion mass $m$, the renormalized coupling $g$, and the renormalization scale $\mu_0$. However the dependence on $g$ and $\mu_0$ is not explicitly shown unless necessary.

The UV divergences in the chiral Ward identity can be smoothed out by cutting the highest eigenvalues of the Dirac operator, by means of a regulator $\R_{\omega_\infty}(\omega) = \exp ( - \omega^2/\omega_\infty^2 )$. The regulated PS correlator and chiral condensate are defined respectively as:
\begin{gather}
C_{PS}^{\R}(x,m,\omega_\infty) = \lim_{L \to \infty} \sum_{a,b} \langle \frac{X_a^\dag(0) X_b(0) X_b^\dag(x) X_a(x)}{(i \omega_a + m)(-i \omega_b + m)} \R_{\omega_\infty} (\omega_a) \R_{\omega_\infty} (\omega_b) \rangle \ ,
\label{eq:reg_pscorr}
\\
\Sigma_{\R}(m,\omega_\infty)=
2m \int_0^\infty \frac{\rho(\omega,m)}{\omega^2 + m^2}  \R_{\omega_\infty}^2(\omega) d\omega \ .
\label{eq:reg_cc}
\end{gather}
When $\omega_\infty$ is sent to infinity and the UV regulator is removed, the regulated $C_{PS}^{\R}(x,m,\omega_\infty)$ becomes the full PS correlator $C_{PS}(x,m)$, while the regulated $\Sigma_{\R}(m,\omega_\infty)$ diverges. The regulated quantities defined above have the nice property of satisfying exactly the chiral Ward identity:
\begin{gather}
\int C_{PS}^{\R}(x,m,\omega_\infty) d^4 x = \frac{\Sigma_\R(m,\omega_\infty)}{m} \ .
\label{eq:reg_ward}
\end{gather}
The reader should refer to appendix~\ref{app:regulated_ward} for the detailed derivation of eq.~\eqref{eq:reg_ward}.

The \textit{second step} consists in splitting and rearranging the integrals over spacetime and Dirac eigenvalues in the regulated chiral Ward identity:
\begin{flalign}
& \int_T^\infty dt \int d^3\mathbf{x} \ C_{PS}^{\R}(x,m,\omega_\infty) - \int_0^{\bar{\omega}} \frac{\rho(\omega,m)}{\omega^2 + m^2}  \R_{\omega_\infty}^2(\omega) d\omega = \nonumber \\
& = -\int_0^T dt \int d^3\mathbf{x} \ C_{PS}^{\R}(x,m,\omega_\infty) + \int_{\bar{\omega}}^\infty \frac{\rho(\omega,m)}{\omega^2 + m^2}  \R_{\omega_\infty}^2(\omega) d\omega \ .
\label{eq:split_ward}
\end{flalign}
This formula segregates the potential leading IR divergences in the lhs and the potential UV divergences in the rhs. However since the lhs is finite when $\omega_\infty$ is sent to infinity, then the UV divergences must cancel out in the rhs as well, and the quantity
\begin{gather}
\Delta W(m,T,\bar{\omega}) = \lim_{\omega_\infty \to \infty} \left\{ - \int_0^T dt \int d^3\mathbf{x} \ C^\R_{PS}(x,m,\omega_\infty) + \int_{\bar{\omega}}^\infty \frac{\rho(\omega,m)}{\omega^2 + m^2}  \R_{\omega_\infty}^2(\omega) d\omega \right\}
\label{eq:DeltaW}
\end{gather}
is finite. Taking the $\omega_\infty \to \infty$ limit of both sides of eq.~\eqref{eq:split_ward} yields:
\begin{gather}
\int_T^\infty dt \int d^3\mathbf{x} \ C_{PS}(x,m) - \frac{\Sigma_{\bar{\omega}}}{2m} = \Delta W(m,T,\bar{\omega}) \ .
\label{eq:master_1}
\end{gather}
where $\Sigma_{\bar{\omega}}$ is the chiral condensate (at nonzero mass) with a sharp UV-regulator:

\begin{gather}
\Sigma_{\bar{\omega}} = 2m \int_0^{\bar{\omega}} \frac{\rho(\omega,m)}{\omega^2 + m^2} d\omega
\ .
\label{eq:sigma_sharp}
\end{gather}
The basic feature of eq.~\eqref{eq:master_1} is the segregation of the potential leading IR divergences (in the $m \to 0$ limit) in the lhs. This is easily seen by inspecting the definition of $\Delta W$ in eq.~\eqref{eq:DeltaW}: \textit{(a)} the integral of the Dirac spectral density is finite in the chiral limit since the region around $\omega=0$ is not included, \textit{(b)} the integral of the regularized PS correlator might be divergent in the chiral limit, but this possible divergence is subleading with respect to the one in the lhs of eq.~\eqref{eq:master_1} because the region around $t=0$ is not included.

\subsection{Spectral decomposition for the PS correlator}

At nonzero fermion mass and in absence of mechanisms that suppress particle decay (like finite volume or large-$N$ limit), any correlator is expected to have a possible pole contribution and a contribution from a continuous density of states with masses above some pair-production threshold. The pole is absent if there is no stable single-particle state in the considered channel. If chiral symmetry is spontaneously broken, even if a small mass is given to the fermions, the PS meson (pion) is the lightest particle of the spectrum, and hence it is stable. In the conformal window, although the PS meson is still the lightest isovector meson (for the Weingarten inequalities~\cite{Weingarten:1983uj}), it might not be the lightest particle with non-trivial isospin. In the latter case, the PS meson might be unstable.

Having said so, the K\"all\'en-Lehmann spectral representation for the PS correlator is
\begin{gather}
C_{PS}(x,m) = \frac{F_{PS}^2 M_{PS}^4}{2 m^2} \Delta(x,M_{PS}^2) + \int_{S_{PS}}^\infty R_{PS}(s,m) \Delta(x,s) ds \ .
\label{eq:ps_spectral}
\end{gather}
$M_{PS}$ is the mass of the pole, and $F_{PS}$ is its electroweak decay constant. The poleless case corresponds to a vanishing decay constant. $S_{PS}$ is the squared threshold energy for the continuum part of the spectrum. $\Delta(x,s)$ is the Euclidean propagator of the free scalar field.

Plugging the spectral decomposition into eq.~\eqref{eq:master_1} one gets:
\begin{gather}
\frac{F_{PS}^2 M_{PS}^2}{2 m^2} e^{- M_{PS} T} + \int_{S_{PS}}^\infty R_{PS}(s,m) e^{- T \sqrt{s}} \frac{ds}{s}  - \frac{\Sigma_{\bar{\omega}}}{2m} = \Delta W(m,T,\bar{\omega}) \ .
\label{eq:master_2}
\end{gather}
I will refer to this equation as the \textit{master equation}. Each term of the master equation will be analyzed separately in sections~\ref{sec:GMOR_large} and~\ref{sec:GMOR_small}. I will refer to the first term in the lhs as the \textit{pole-term}, to the second one as the \textit{$R$-term}, to the third one as the \textit{$\Sigma$-term}, and finally to the rhs as the \textit{reminder}.

\section{RG analysis}
\label{sec:RG}

\subsection{General formulae}
\label{sec:RG:general}

The renormalization procedure implies the introduction of a renormalization scale $\mu_0$ and the definition of suitable renormalized parameters: the renormalized coupling constant $g$ and the renormalized mass $m$. The choice of $\mu_0$ is arbitrary and has no particular physical meaning. In all the physical observables, a change of the scale $\mu_0 \to \mu$ can be reabsorbed by changing the parameters of the theory $g\to \bar{g}(\mu)$ and $m \to \bar{m}(\mu)$. The running parameters $\bar{g}(\mu)$ and $\bar{m}(\mu)$ are governed by the RG-equations, and can be traded for the RG-invariant parameters $\Lambda$ and $M$. IR conformality is modeled with an IR fixed point in the coupling constant $g(0)=g_*$. The reader is referred to appendix~\ref{app:RG} for an overview of the notations.

Let be $O(E)$ an observable with the following properties: \textit{(a)} it renormalizes multiplicatively; \textit{(b)} it is a function of a physical (i.e. RG-invariant) energy $E$. As shown in appendix~\ref{app:RG}, one can solve the RG-equation and use the existence of an IR-fixed point to obtain the general form:
\begin{gather}
O(E,g,m,\mu_0) =
Z_O \left(\frac{\mu_0}{\Lambda} \right) \mu_0^{\gamma_*^{(O)}} E^{d_O-\gamma_*^{(O)}}  \hat{O} \left( \frac{M}{E} , \frac{E}{\Lambda} \right) \ ,
\label{eq:RG_O_5}
\end{gather}
where $Z_O(\mu_0/\Lambda)$ is a renormalization factor, and $\gamma_*^{(O)}$ is the anomalous dimension of $O$ at the IR-fixed point.

The RG-analysis for the mass or the decay constant of a particle is much simpler. These are RG-invariant quantities, hence they do not depend on the RG-scale $\mu_0$ once everything is expressed in terms of $\Lambda$ and $M$. If $M_X$ is one of those observables, its general form is:
\begin{gather}
M_X(g,m,\mu_0) = M \hat{m}_X \left( \frac{M}{\Lambda} \right) \ .
\label{eq:RG_MX}
\end{gather}

\medskip

At this point one would like to sit in the chiral point $M=0$ and see from eq.~\eqref{eq:RG_O_5} that the observable $O$ exhibits the power law at small energies that is typical of IR-conformal theories. However this cannot be inferred unless one assumes that the limits $\Lambda \to \infty$ and $M \to 0$ are regular. Various combination of these limits identify different regimes of the theory, which I will briefly describe separately.

\begin{itemize}

\item \textit{Conformal theory}: $M \to 0$ and $\Lambda \to \infty$. Notice that the $\Lambda \to \infty$ is the same as the limit $g \to g_*$ at fixed renormalization scale $\mu_0$ (figure~\ref{fig1}). Asymptotic freedom is washed away and the theory becomes exactly scale invariant. Assuming regularity of the RG-invariant functions $\hat{m}_X$ and $\hat{O}$, one obtains that all masses vanish and the observable $O$ behaves like an exact power law in the energy scale $E$:
\begin{gather}
M_X = 0 \ ,
\label{eq:RG_MX_CT}
\\
O(E) =
\mu_0^{\gamma_*^{(O)}} E^{d_O-\gamma_*^{(O)}} \hat{O}(0,0) \ .
\label{eq:RG_O_CT}
\end{gather}

\begin{figure}
\includegraphics[width=.6\textwidth]{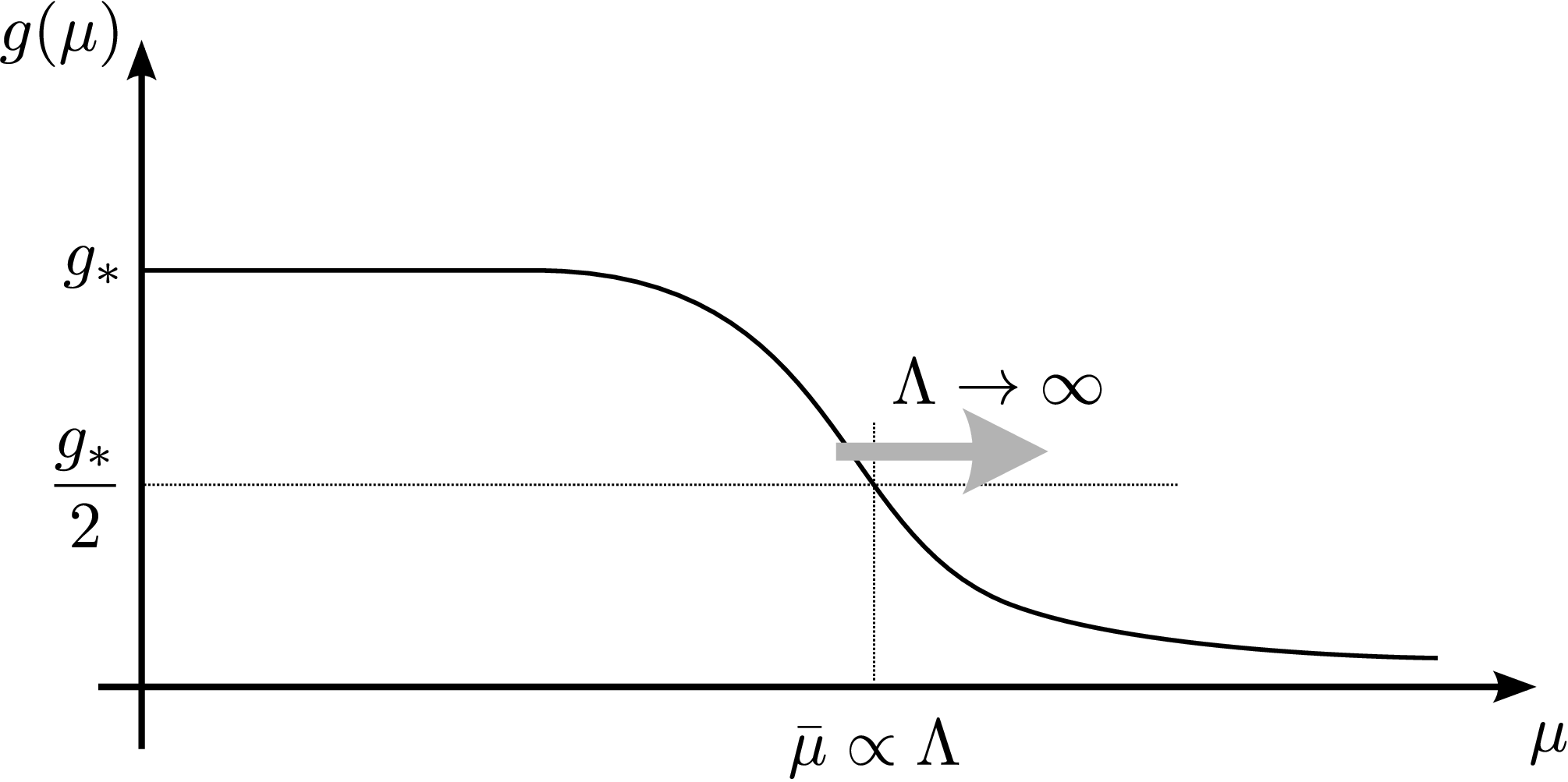}
\caption{\label{fig1}
Sketch of the running coupling for an IR-conformal UV-free theory. The transition between the IR-conformal region (at small $\mu$) and the asymptotic-freedom region (at large $\mu$) happens at an energy scale which is proportional to $\Lambda$. For instance the transition energy can defined as the value of $\bar{\mu}$ at which the running coupling is half of its value at the fixed point. If $\Lambda$ is sent to infinity, also $\bar{\mu}$ goes to infinity, the asymptotic freedom is washed away, and the running coupling becomes $g(\mu)=g_*$. In this limit the theory is exactly scale invariant.
}
\end{figure}

\item \textit{Mass-deformed UV-conformal theory}: $\Lambda \to \infty$ and $M$ finite. As before, the running coupling is tuned at the value of the fixed point, but scale invariance is broken by the mass:
\begin{gather}
M_X = M \hat{m}_X(0) \ ,
\label{eq:RG_MX_mCT}
\\
O(E) =
\mu_0^{\gamma_*^{(O)}} E^{d_O-\gamma_*^{(O)}}  \hat{O} \left( \frac{M}{E} , 0 \right) \ .
\label{eq:RG_O_mCT}
\end{gather}
Particles stay massive and all the masses are proportional to the RG-invariant fermion mass $M$. Scale invariance is approximatively restored at small distances/high energies, in fact the observables $O$ behaves like a power law asymptotically in the regime $E \gg M$. Notice that if $\Lambda$ is set to infinity, the high energy limit is formally identical to the chiral limit (only the ratio $E/M$ matters).

\item \textit{IR-conformal theory}: $M \to 0$ and $\Lambda$ finite:
\begin{gather}
M_X = 0 \ ,
\label{eq:RG_MX_IRCT}
\\
O(E) =
Z_O \left(\frac{\mu_0}{\Lambda} \right) \mu_0^{\gamma_*^{(O)}} E^{d_O-\gamma_*^{(O)}}  \hat{O} \left( 0 , \frac{E}{\Lambda} \right) \ .
\label{eq:RG_O_IRCT}
\end{gather}
In this limit the theory is asymptotically free and exhibits asymptotic scale invariance at large distances/low energies. Masses vanish, while the observable $O$ exhibits the power-law behavior described in eq.~\eqref{eq:RG_O_CT} asymptotically at low energies $E \ll \Lambda$. Notice that if $M$ is set to zero, the low energy limit is formally identical to the $\Lambda \to \infty$ limit (only the ratio $E/\Lambda$ matters).

\end{itemize}

Summarizing, eqs.~\eqref{eq:RG_MX_CT}, \eqref{eq:RG_O_CT}, \eqref{eq:RG_MX_mCT}, \eqref{eq:RG_O_mCT}, \eqref{eq:RG_MX_IRCT} and \eqref{eq:RG_O_IRCT} are obtained from the RG equations in presence of an IR fixed point, under the assumptions that the beta function and the anomalous dimensions are regular at the fixed point, and that the UV-conformal ($\Lambda \to \infty$) and IR-conformal ($M \to 0$) limits of the observables $M_X$ and $O(E)$ are finite. I will refer to this set of assumptions as the \textit{hyperscaling hypothesis}.

\subsection{Specialization to relevant observables}

I want to specialize now the general formulae obtained in the previous subsection to the observables we are interested in. The PS mass $M_{PS}$, decay constant $F_{PS}$ and square root of the threshold energy $S_{PS}$ are RG-invariant quantities with dimension of a mass. Therefore the general formula~\eqref{eq:RG_MX} applies:
\begin{gather}
M_{PS} = M \hat{m}_{PS}(M/\Lambda) \ ,
\label{eq:rg_mps}
\\
F_{PS} = M \hat{f}_{PS}(M/\Lambda) \ ,
\label{eq:rg_fps}
\\
S_{PS} = M^2 \hat{s}_{PS}(M/\Lambda) \ .
\label{eq:rg_sps}
\end{gather}

The relation between the fermion mass $m$ at the scale $\mu_0$ and the RG-invariant mass $M$ is given by (see appendix~\ref{app:RG}):
\begin{equation}
m = 
\frac{Z(M/\Lambda)}{Z(\mu_0/\Lambda)} \mu_0^{-\gamma_*} M^{1+\gamma_*}
\ .
\label{eq:rgi_m}
\end{equation}

From the Ward identities for the chiral symmetry, one can easily compute the anomalous dimensions of the PS correlator ($\gamma_{C_{PS}} = 2 \gamma$), the PS spectral density ($\gamma_{R_{PS}} = 2 \gamma$), and the Dirac eigenvalue density ($\gamma_{\rho} = \gamma$). Specializing the general formula~\eqref{eq:RG_O_5} to the PS correlator (with $E=|x|^{-1}$ and $d_O=6$), and to the PS spectral density (with $E=\sqrt{s}$ and $d_O=2$) is straightforward:
\begin{gather}
C_{PS}(x) =
Z^2(\mu_0/\Lambda) \mu_0^{2\gamma_*} |x|^{-6+2\gamma_*}
\ \hat{C}_{PS} \left( M |x|, \frac{1}{\Lambda |x|} \right) \ ,
\label{eq:rg_cps} \\
R_{PS}(s) =
Z^2(\mu_0/\Lambda) \mu_0^{2\gamma_*} s^{1-\gamma_*}
\ \hat{R}_{PS} \left( \frac{M}{\sqrt{s}}, \frac{\sqrt{s}}{\Lambda} \right) \ .
\label{eq:rg_rps}
\end{gather}

Specializing the general formula~\eqref{eq:RG_O_5} to the Dirac eigenvalue density is slightly more involved, because of the dependence on the parameter $\omega$ that is not RG-invariant. Since the eigenvalue has the same anomalous dimension as the mass, one can use a relation similar to eq.~\eqref{eq:rgi_m} in order to define a RG-invariant eigenvalue $\Omega$:
\begin{equation}
\omega = 
\frac{Z(\Omega/\Lambda)}{Z(\mu_0/\Lambda)} \mu_0^{-\gamma_*} \Omega^{1+\gamma_*}
\ .
\label{eq:rgi_omega}
\end{equation}
Finally one can take eq.~\eqref{eq:RG_O_5} with $E=\Omega$ and $d_O=3$ and get:
\begin{gather}
\rho(\omega) =
Z(\mu_0/\Lambda) \mu_0^{\gamma_*} \Omega^{3-\gamma_*}
\ \hat{\rho} \left( \frac{M}{\Omega}, \frac{\Omega}{\Lambda} \right) \ .
\label{eq:rg_rho}
\end{gather}

\section{GMOR-like relation for large anomalous dimensions}
\label{sec:GMOR_large}

Since the PS correlator falls off like $|x|^{-6+2\gamma_*}$ in the chiral limit, the PS susceptibility is IR-convergent in the same limit if $0<\gamma_*<1$, while it is divergent for $1\le \gamma_*<2$.\footnote{
The case $\gamma_*=1$ gives rise to logarithmic divergences in the PS susceptibility, and would require a separate mathematical analysis. I will just ignore this marginal case for the rest of the paper.
} In this section I will show that all the terms in the lhs of the master equation~\eqref{eq:master_2} diverge like $M^{2-2\gamma_*}$ in the chiral limit. Since the reminder of the master equation~\eqref{eq:master_2} is subleading in the chiral limit, the divergences in the lhs must cancel each other. This condition will lead to the GMOR-like relation.

\subsection{Chiral limit of the master equation}
\label{sec:GMOR_large:master}

Let us focus on the pole term of the master equation~\eqref{eq:master_2}. Close to the chiral limit, the relation~\eqref{eq:rgi_m} between the running mass $m$ and the RG-invariant mass $M$ becomes:
\begin{gather}
m \simeq Z_0^{-1} \mu_0^{-\gamma_*} M^{1+\gamma_*} \ ,
\end{gather}
where I have introduced the shorthand notation $Z_0=Z(\mu_0/\Lambda)$.

Using the general formulae \eqref{eq:rg_mps} and \eqref{eq:rg_fps} obtained in the previous section, the leading behavior of the pole-term in the chiral limit is extracted:
\begin{gather}
\frac{F_{PS}^2 M_{PS}^2}{2 m^2} e^{- M_{PS} T} \simeq
\frac{1}{2} Z^2_0 \hat{f}_{PS}^2 \hat{m}_{PS}^2 \mu_0^{2\gamma_*} M^{2-2\gamma_*} \ .
\label{eq:chiral_pole}
\end{gather}
Both the functions $\hat{f}_{PS}$ and $\hat{m}_{PS}$ with no argument have to be meant evaluated in zero.

\medskip

Let us focus now on the $R$-term of the master equation~\eqref{eq:master_2}. Since $R_{PS}(s,0) \propto s^{1-\gamma_*}$ for small values of the squared energy $s$ in the chiral limit, the integrand goes like $s^{-\gamma_*}$. In the range of anomalous dimensions $1<\gamma_*<2$ the integral is divergent. In order to single out the divergence, one has to plug formula~\eqref{eq:rg_rps} into the R-term of the the master equation~\eqref{eq:master_2}, and to change the integration variable accordingly to $M^2=x^2 s$. One gets:
\begin{gather}
\int_{S_{PS}}^\infty R_{PS}(s,m) e^{- T \sqrt{s}} \frac{ds}{s} \simeq
2 Z^2_0 \mu_0^{2\gamma_*} M^{2-2\gamma_*}
\int_0^{\hat{s}_{PS}^{-1/2}} x^{-3+2\gamma_*} \hat{R}_{PS} (x,0) dx \ .
\label{eq:chiral_R}
\end{gather}
Again, the function $\hat{s}_{PS}$ with no argument has to be understood as evaluated in zero. Notice that the integral in the rhs is finite because $\hat{R}_{PS} (x,0)$ is finite at $x=0$ (for hyperscaling hypothesis). Moreover the leading term in the mass does not depend on the UV regulator $T^{-1}$ and on the scale $\Lambda$ of asymptotic freedom.

\medskip

Consider finally the $\Sigma$-term of the master equation~\eqref{eq:master_2}. Since $\rho(\omega,0) \propto \omega^{\frac{3-\gamma_*}{1+\gamma_*}}$ for small values of the eigenvalue in the chiral limit, the integrand goes like $\omega^{\frac{1-3\gamma_*}{1+\gamma_*}}$. In the range of anomalous dimensions $1<\gamma_*<2$, the exponent $\frac{1-3\gamma_*}{1+\gamma_*}$ is smaller than $-1$ giving rise to a divergent integral. In order to single out the divergence, one has to plug formula~\eqref{eq:rg_rps} and to change the integration variable accordingly to $M=x\Omega$, where $\Omega$ is implicitly defined in terms of $\omega$ in eq.~\eqref{eq:rgi_omega}. After some long but straightforward manipulations, one gets:
\begin{gather}
\frac{\Sigma_{\bar{\omega}}}{2m} \simeq
Z^2_0 \mu_0^{2\gamma_*} M^{2-2\gamma_*} (1 + \gamma_*)
\int_0^\infty
\frac{x^{-3+2\gamma_*} \hat{\rho}(x,0)}{1+x^{2+2\gamma_*}} dx \ .
\label{eq:chiral_Sigma}
\end{gather}
The integral in the rhs is convergent because $\hat{\rho}(0,0)$ is finite (for hyperscaling hypothesis). Notice that the leading term in the mass does not depend on the UV regulator $\bar{\omega}$ and on the scale $\Lambda$ of asymptotic freedom.

\subsection{GMOR-like relation}
\label{sec:GMOR_large:GMOR}

Summing up, all the terms in the lhs of the master equation are equally IR-divergent (like $M^{2-2\gamma_*}$) in the chiral limit, if the anomalous dimension is in the range $1 < \gamma_* < 2$, while the remainder of the master equation is subleading in the chiral limit. A GMOR-like relation is obtained by discarding the rhs of the master equation and dividing by the pole contribution:
\begin{gather}
\lim_{m \to 0} \frac{m \Sigma_{\bar{\omega}}(m)}{F_{PS}^2 M_{PS}^2} = 1 +
%\lim_{m \to 0} \frac{2m^2}{F_{PS}^2 M_{PS}^2} \int_{S_{PS}}^{S_\infty} R_{PS}(s,m) \frac{ds}{s} \ .
\frac{4}{\hat{f}_{PS}^2 \hat{m}_{PS}^2} \int_0^{\hat{s}_{PS}^{-1/2}} x^{-3+2\gamma_*} \hat{R}_{PS} (x,0) dx \ .
\label{eq:gmor_large_gamma}
\end{gather}
Eq.~\eqref{eq:gmor_large_gamma} is the direct generalization of the GMOR relation to the IR-conformal theory with anomalous dimension in the range $1 < \gamma_* < 2$. The pole dominance is not realized in this case: the continuous part of the PS spectrum has the same weight as the pole in the regulated Ward identity~\eqref{eq:reg_ward}. Notice that the continuous part of the spectrum gives always a positive contribution to the rhs of eq.~\eqref{eq:gmor_large_gamma}.

The GMOR-like relation~\eqref{eq:gmor_large_gamma} does not depend on $\Lambda$, and therefore does not distinguish between the asymptotically free and UV-conformal theories. In the case of the (mass-deformed) UV-conformal theory, the GMOR-like relation~\eqref{eq:gmor_large_gamma} can be obtained much more easily. In fact at infinite $\Lambda$ the PS correlator becomes UV integrable since it goes like $x^{6-2\gamma_*}$ for small values of $x$ ($2 < 6-2\gamma_* < 4$). Also the chiral condensate becomes UV finite since the Dirac eigenvalue density goes like $\omega^{\frac{3-\gamma_*}{1+\gamma_*}}$ for large eigenvalues $\omega$ ($\frac{1}{3} < \frac{3-\gamma_*}{1+\gamma_*} < 1$). Therefore the UV cutoff $\omega_\infty$ can be safely removed in the Ward identity~\eqref{eq:reg_ward}, and eq.~\eqref{eq:gmor_large_gamma} can be shown to be valid at every mass:
\begin{gather}
\frac{m \Sigma(m)}{F_{PS}^2 M_{PS}^2} = 1 +
\frac{2m^2}{F_{PS}^2 M_{PS}^2} \int_{S_{PS}}^{\infty} R_{PS}(s,m) \frac{ds}{s} \ .
\end{gather}

\section{GMOR-like relation for small anomalous dimensions}
\label{sec:GMOR_small}

\subsection{$1/3 < \gamma_* < 1$}
\label{sec:GMOR_small_1}

For values of the anomalous dimension in the range $0<\gamma_*<1$, the terms in the lhs of the master equation~\eqref{eq:master_2} do not contain IR-divergences. The chiral limit of the regulated Ward identity~\eqref{eq:reg_ward} yields:
\begin{gather}
\int C_{PS}^{\R}(x,0,\omega_\infty) d^4 x = \int_0^\infty \frac{\rho(\omega,0)}{\omega^2}  \R_{\omega_\infty}^2(\omega) d\omega \ .
\end{gather}
In fact the PS correlator decays like $|x|^{-6+2\gamma_*}$ at large distances, which is integrable for $0<\gamma_*<1$. While the integrand in the rhs vanishes like $\omega^{\frac{1-3\gamma_*}{1+\gamma_*}}$ at small eigenvalues, which is also integrable for $0<\gamma_*<1$. The argument used in section~\ref{sec:GMOR_large} for large anomalous dimensions does not hold anymore. However one can take derivatives of the Ward identity~\eqref{eq:ward} with respect to the mass. Taking for instance the first derivative, one gets eq.~\eqref{eq:ward_first}. I will show that both sides of eq.~\eqref{eq:ward_first} are IR divergent if the anomalous dimension is in the range $1/3 < \gamma_* < 1$.\footnote{
Also the case $\gamma_*=1/3$ gives rise to IR-divergences in both sides of eq.~\eqref{eq:ward_first}. However these divergences are logarithmic and would require a separate mathematical analysis. I will just ignore this marginal case in the rest of the paper.
} Matching the IR-divergences in the two sides of eq.~\eqref{eq:ward_first} will yield the generalized GMOR relation in this case. Of course, in order to make everything well defined, one should introduce a regulated version of eq.~\eqref{eq:ward_first}, by taking the first derivative with respect the mass of the master equation in the form~\eqref{eq:master_1}:
\begin{gather}
\int_T^\infty dt \int d^3\mathbf{x} \int d^4y \
\langle \bar{\psi}\psi(y) P^\dag(x) P(0) \rangle_{c,m} -
\frac{\partial}{\partial m} \left[ \frac{\Sigma_{\bar{\omega}}}{2m} \right] = \frac{\partial \Delta W}{\partial m}(m,T,\bar{\omega}) \ .
\label{eq:master_1_1der}
\end{gather}
In complete analogy with the analysis of the reminder of the master equation~\eqref{eq:master_2}, if any IR divergence is present, the derivative $\partial \Delta W/\partial m$ contains only subleading divergences. Matching the leading divergences in the rhs of the last equation, I will obtain a generalized GMOR relation for $1/3 < \gamma_* < 1$.

The first step consists in identifying the leading divergence in the mass of the two terms in the lhs of eq.~\eqref{eq:master_1_1der}. The RG-analysis leads for the 3-point function to the following functional form:
\begin{gather}
\langle \bar{\psi}\psi(y) P^\dag(x) P(0) \rangle_{c,m} = \frac{Z_0^3 \mu_0^{3\gamma_*}}{|x|^{3-\gamma*} |y|^{3-\gamma*} |x-y|^{3-\gamma*}} \hat{C}_3 \left( \frac{xy}{|x||y|} , M|x|, M|y|, \frac{M}{\Lambda} \right) \ ,
\end{gather}
where the function $\hat{C}_3$ is finite in the chiral limit and also in the UV-conformal ($\Lambda \to \infty$) limit, and it is exponentially decaying at large distances for nonvanishing mass. The coordinate dependence in front of the function $\hat{C}_3$ comes from the assumption that the theory at $\Lambda = \infty$ is invariant under the full conformal group (and not only under scale transformations).

Integrating the 3-point function in $x$ and $y$, and rescaling the integration variables with $M$, one gets the following chiral behavior:
\begin{flalign}
& \int_T^{\infty} dt \int d^3\mathbf{x} \int d^4y \langle \bar{\psi}\psi(y) P^\dag(x) P(0) \rangle_c \simeq \nonumber \\
& \simeq Z_0^3 \mu_0^{3\gamma_*} M^{1-3\gamma_*}
\int d^4x \int d^4\hat{y} \frac{\hat{C}_3 \left( \frac{\hat{x}\hat{y}}{|\hat{x}||\hat{y}|} , |\hat{x}|, |\hat{y}|, 0 \right)}{|\hat{x}|^{3-\gamma*} |\hat{y}|^{3-\gamma*} |\hat{x}-\hat{y}|^{3-\gamma*}} \ .
\label{eq:chiral_3point}
\end{flalign}
For $1/3 < \gamma_* < 1$ the integral is finite and the whole quantity diverges like $M^{1-3\gamma_*}$. Since the rhs of eq.~\eqref{eq:master_1_1der} is finite in the chiral limit, also the second term of its lhs must be divergent like $M^{1-3\gamma_*}$, in such a way that the two divergences cancel each other:
\begin{flalign}
\frac{\partial}{\partial m} \int_T^\infty dt \int d^3\mathbf{x} C_{PS}(x,m) & = 
\int_T^\infty dt \int d^3\mathbf{x} \int d^4y \
\langle \bar{\psi}\psi(y) P^\dag(x) P(0) \rangle_{c,m} \simeq \nonumber \\
& \simeq \frac{\partial}{\partial m} \left[ \frac{\Sigma_{\bar{\omega}}}{2m} \right] = O(M^{1-3\gamma_*})
\ .
\end{flalign}
At the leading order one can also replace the derivatives with the incremental ratios:
\begin{gather}
\frac{1}{m} \int_T^\infty dt \int d^3\mathbf{x} [C_{PS}]_{sub}(x,m) \simeq
\frac{1}{m} \left[ \frac{\Sigma_{\bar{\omega}}}{2m} \right]_{sub} = O(M^{1-3\gamma_*})
\ ,
\label{eq:almost_gmor_small}
\end{gather}
where I use the shorthand notation $[F]_{sub}(m) = F(m)-F(0)$. Using the spectral decomposition for the PS correlator in eq.~\eqref{eq:ps_spectral}, and dividing all terms in eq.~\eqref{eq:almost_gmor_small} by the pole contribution, one readily obtains:
\begin{gather}
\lim_{m \to 0} \frac{2 m^2}{F_{PS}^2 M_{PS}^2} \left[ \frac{\Sigma_{\bar{\omega}}}{2m} \right]_{sub} %= 1 + \lim_{m \to 0} \frac{2 m^2}{F_{PS}^2 M_{PS}^2} \int_{S_{PS}}^{S_\infty} [R_{PS}]_{sub}(s,m) \frac{ds}{s} \ ,
=1 + \frac{4}{\hat{f}_{PS}^2 \hat{m}_{PS}^2} \int_0^{\hat{s}_{PS}^{-1/2}} x^{-3+2\gamma_*} [\hat{R}_{PS} (x,0) - \hat{R}_{PS}(0,0)] dx \ ,
\label{eq:gmor_small}
\end{gather}
which is the generalized GMOR relation for anomalous dimensions in the range $1/3 < \gamma_* < 1$.

\subsection{$0 < \gamma_* < 1/3$}
\label{sec:GMOR_small_2}

If the anomalous dimension is in the range $0 < \gamma_* < 1/3$, then also the first derivative~\eqref{eq:master_1_1der} of the master equation does not contain IR divergences. One has to consider the second derivative with respect to the mass:
\begin{gather}
\int_T^\infty dt \int d^3\mathbf{x} \int d^4yd^4z \
\langle \bar{\psi}\psi(y) \bar{\psi}\psi(z) P^\dag(x) P(0) \rangle_{c,m} -
\frac{\partial^2}{\partial m^2} \left[ \frac{\Sigma_{\bar{\omega}}}{2m} \right] = \frac{\partial^2 \Delta W}{\partial m^2}(m,T,\bar{\omega}) \ .
\label{eq:master_1_2der}
\end{gather}
If $\Delta x$ is the minimum distance between any of the pairs of operators in the 4-point function in the lhs, for large $\Delta x$ the 4-point function in the chiral limit falls off like $(\Delta x)^{-12+4\gamma_*}$, which makes it non integrable in the IR for any positive anomalous dimension. Repeating the RG-analysis for the 4-point function one can show that the first term in the lhs of eq.~\eqref{eq:master_1_2der} diverges like $M^{-4\gamma_*}$ in the chiral limit. Again, the rhs of eq.~\eqref{eq:master_1_2der} is IR-subleading and it can be discarded in the chiral limit. The divergence of the first term in the lhs must be canceled by an equal divergence in the second term:
\begin{flalign}
\frac{\partial^2}{\partial m^2} \int_T^\infty dt \int d^3\mathbf{x} C_{PS}(x,m) & = 
\int_T^\infty dt \int d^3\mathbf{x} \int d^4yd^4z \
\langle \bar{\psi}\psi(y) \bar{\psi}\psi(z) P^\dag(x) P(0) \rangle_{c,m} \simeq \nonumber \\
& \simeq \frac{\partial^2}{\partial m^2} \left[ \frac{\Sigma_{\bar{\omega}}}{2m} \right] = O(M^{-4\gamma_*})
\ .
\label{eq:almost_gmor_small_2}
\end{flalign}

I want to replace now the derivatives with finite differences. For instance one can Taylor-expand the the PS susceptibility, and write the first derivative in terms of the connected 3-point function:
\begin{flalign}
& \frac{\partial^2}{\partial m^2} \int_T^\infty dt \int d^3\mathbf{x} C_{PS}(x,m) \simeq \nonumber \\
& \simeq \frac{1}{m^2} \int_T^\infty dt \int d^3\mathbf{x} [C_{PS}]_{sub}(x,m) - \frac{1}{m} \int_T^\infty dt \int d^3\mathbf{x} \int d^4y \langle \bar{\psi}\psi(y) P^\dag(x) P(0) \rangle_{c,m=0} \ .
\end{flalign}
However the 3-point function can be constrained by the non-anomalous subgroup $\mathbf{Z}_{4 T_R N_f}$ ($R$ is the representation of the fermions under gauge transformations) of the axial symmetry. In the conformal window the axial $\mathbf{Z}_{4 T_R N_f}$ is expected not to be broken along with the chiral symmetry. If $T_R N_f$ is an integer\footnote{This happens for instance for an even number of fundamental or (anti)symmetric two-index fermions, or for any number of flavors of adjoint (Dirac) fermions.}, then the transformation $\psi \to i \gamma_5 \psi$ is a symmetry in the chiral limit. The 3-point function $\langle \bar{\psi}\psi(y) P^\dag(x) P(0) \rangle_{c,m=0}$ is odd under this symmetry, and therefore it must be zero. The previous equation reduces to:
\begin{gather}
\frac{\partial^2}{\partial m^2} \int_T^\infty dt \int d^3\mathbf{x} C_{PS}(x,m) \simeq
\frac{1}{m^2} \int_T^\infty dt \int d^3\mathbf{x} [C_{PS}]_{sub}(x,m) \ .
\end{gather}
For a nonvanishing fermion mass the transformation $\psi \to i \gamma_5 \psi$ is still a symmetry if combined with $m \to -m$. This gives for instance that the Dirac eigenvalue density is even under mass-sign flip, while the regulated chiral condensate $\Sigma_{\bar{\omega}}$ is odd. This yields for the second derivative of $\Sigma_{\bar{\omega}}/(2m)$ the following asymptotic equality:
\begin{gather}
\frac{\partial^2}{\partial m^2} \left[ \frac{\Sigma_{\bar{\omega}}}{2m} \right] \simeq \frac{1}{m^2} \left[ \frac{\Sigma_{\bar{\omega}}}{2m} \right]_{sub} \ ,
\end{gather}
since the first derivative vanishes in the chiral limit. The condition~\eqref{eq:almost_gmor_small_2} for the divergences to cancel each other can be hence rewritten by replacing the derivatives with finite differences:
\begin{gather}
\frac{1}{m^2} \int_T^\infty dt \int d^3\mathbf{x} [C_{PS}]_{sub}(x,m) \simeq
\frac{1}{m^2} \left[ \frac{\Sigma_{\bar{\omega}}}{2m} \right]_{sub} = O(M^{-4\gamma_*})
\ ,
\end{gather}
which is completely equivalent to the corresponding eq.~\eqref{eq:almost_gmor_small} for anomalous dimensions in the range $1/3<\gamma_*<1$. Therefore one gets for $0<\gamma_*<1/3$ the same generalized GMOR equation~\eqref{eq:gmor_small} as for the $1/3<\gamma_*<1$ case.

\section{GMOR ratio on the lattice}
\label{sec:lattice}

In the past few years, many lattice investigations have been devoted to the study of the conformal window. It is natural to ask whether the generalized GMOR relations can be used in lattice simulations to discriminate IR-conformality from chiral symmetry breaking.

In case of IR-conformality, the two GMOR relations~\eqref{eq:gmor_large_gamma} and \eqref{eq:gmor_small} involve the PS spectral density, which is not easy to be computed on the lattice. However one can focus on the ratio $m \Sigma_{\bar{\omega}} /(F_{PS}^2 M_{PS}^2)$ (that I will call \textit{GMOR ratio}). I have shown in sects.~\ref{sec:GMOR_large} and \ref{sec:GMOR_small} that the GMOR ratio is always greater or equal to one, and its value discriminates between the following three situations:
\begin{gather}
\begin{cases}
\lim_{m \to 0} \frac{m \Sigma_{\bar{\omega}}(m)}{F_{PS}^2 M_{PS}^2} = 1 & \qquad \textrm{for spontaneously-broken chiral symmetry,} \\
\lim_{m \to 0} \frac{m \Sigma_{\bar{\omega}}(m)}{F_{PS}^2 M_{PS}^2} = \infty & \qquad \textrm{for IR-conformality with $0 < \gamma_* < 1$,} \\
1< \lim_{m \to 0} \frac{m \Sigma_{\bar{\omega}}(m)}{F_{PS}^2 M_{PS}^2} < \infty & \qquad \textrm{for IR-conformality with $1 < \gamma_* < 2$.}
\end{cases}
\end{gather}
I remind that the chiral condensate $\Sigma_{\bar{\omega}}(m)$ appearing in the GMOR ratio is regularized with a renormalized sharp cutoff $\bar{\omega}$ in the eigenvalue space, and hence it is a well-defined finite quantity in the continuum theory at any mass.

Let us focus now on the lattice-discretized theory. All the dimensionless quantities defined in the lattice theory will be denoted with a hat. For definiteness I will consider Wilson fermions.\footnote{
The GMOR relation with Wilson fermions is already discussed in~\cite{Giusti:1998wy} in the case of broken chiral symmetry.
} Since the Wilson-Dirac operator $\hat{D}_W$ is in general nondiagonalizable, it is useful to introduce the positive operator $\hat{H}_W = \hat{D}_W^\dag \hat{D}_W$. Let $\hat{\mathbb{P}}(\hat{\omega})$ be the projector on the eigenspaces of $\hat{H}_W$ corresponding to eigenvalues smaller than $\hat{m}^2+\hat{\omega}^2$, being $\hat{m}$ the quark mass defined via the PCAC relation. The discretized version of $\Sigma_{\bar{\omega}}(m)$ is:
\begin{gather}
\hat{\Sigma}(\hat{g},\hat{m},\hat{\omega}) = \frac{\hat{m}}{\hat{V}} \langle \tr \frac{\hat{\mathbb{P}}(\hat{\omega})}{\hat{H}_W} \rangle_{g,\hat{m}} \ ,
\end{gather}
where $\hat{g}$ is the bare coupling constant, and $\hat{V}$ is the volume in lattice units. The discretized chiral condensate $\hat{\Sigma}(\hat{g},\hat{m},\hat{\omega})$ is computable on the lattice (for an implementation of the projector $\hat{\mathbb{P}}(\hat{\omega})$, see~\cite{Giusti:2008vb}). How do you take the continuum limit? One needs a dimensionful quantity to set the scale (for instance the Sommer radius $r_0$)\footnote{
The Sommer radius $r_0$~\cite{Sommer:1993ce} is defined from the static potential $V(r)$ as
\begin{gather}
r_0^2 \frac{dV}{dr}(r_0) = c \ .
\end{gather}
If the constant $c$ is suitably chosen, the Sommer radius is well defined and finite also for an IR-conformal theory.
}, and the PS mass $M_{PS}$ to set the fermion mass. If $\hat{r}_0$ and $\hat{M}_{PS}$ are respectively the Sommer radius and the PS mass in the discretized theory, by requiring that:
\begin{gather}
\hat{r_0} \hat{M}_{PS}(\hat{g},\hat{m}) = r_0 M_{PS} \ ,
\end{gather}
one selects a trajectory $\hat{m}(\hat{g})$ in the parameter space. This trajectory defines how you have to change the bare finite mass $\hat{m}$ while going to the continuum limit ($\hat{g} \to 0$) in order to keep the physical (renormalized) fermion mass finite.

Before taking the continuum limit, one has to give also a similar prescription for the eigenvalue cutoff $\hat{\omega}$. One can consider for instance the mode number per unit volume:
\begin{gather}
\hat{\zeta}(\hat{g},\hat{m},\hat{\omega}) = \frac{1}{\hat{V}} \langle \tr \hat{\mathbb{P}}(\hat{\omega}) \rangle_{\hat{g},\hat{m}} \ .
\end{gather}
The authors of~\cite{Giusti:2008vb} showed that this is a RG-invariant quantity also with Wilson fermions. Once $\hat{\omega}$ is renormalized, the continuum limit can be taken without extra renormalization factors. One can reverse the argument in order to define the renormalization of $\hat{\omega}$. Choose a value $\bar{\zeta}$ for the mode number per unit volume in the continuum. By requiring that the dimensionless combination $r_0^4 \bar{\zeta}$ is fixed at any finite lattice spacing (and therefore in the continuum limit):
\begin{gather}
\hat{r}_0^4 \hat{\zeta}(\hat{g},\hat{m}(g),\hat{\omega}) = r_0^4 \bar{\zeta} \ ,
\label{eq:modenumber}
\end{gather}
one selects a trajectory $\hat{\omega}(\hat{g})$. This trajectory defines how you have to change the bare eigenvalue cutoff $\hat{\omega}$ while going to the continuum limit ($\hat{g} \to 0$) in order to keep finite the mode number per unit volume in physical units.

The continuum limit of the GMOR ratio is readily taken:
\begin{gather}
\lim_{\hat{g} \to 0} \frac{\hat{m}(g) \hat{\Sigma}(\hat{g},\hat{m}(\hat{g}),\hat{\omega}(\hat{g}))}{\hat{F}_{PS}^2 \hat{M}_{PS}^2(\hat{g},\hat{m}(g))} = \frac{m \Sigma_{\bar{\omega}(\bar{\zeta})}(m)}{F_{PS}^2 M_{PS}^2} \ .
\end{gather}
This procedure selects implicitly a value for the renormalized cutoff $\bar{\omega}$ that depends on the chosen value of $\bar{\zeta}$. The relation between $\bar{\omega}$ and $\bar{\zeta}$ can be written by taking the continuum limit of~\eqref{eq:modenumber} and writing the mode number per unit volume in terms of the eigenvalue density distribution:
\begin{gather}
\int^{\bar{\omega}}_{-\bar{\omega}} \rho(\omega,m) d\omega = \bar{\zeta} \ .
\end{gather}
However we are not interested in the particular value of $\bar{\omega}$, since I have already shown that the GMOR ratio does not depend on it in the chiral limit.

Summarizing, the GMOR ratio is a well-defined quantity and can be defined also on the lattice (even with a nonchiral formulation of the fermions). Its behaviour in the chiral limit allows to discriminate between chiral symmetry breaking, IR-conformality with small anomalous dimensions, and IR-conformality with large anoumalous dimensions.

\section{Conclusions}

The well-known GMOR relation for QCD-like theories that break chiral symmetry spontaneously
\begin{gather}
\lim_{m \to 0} \frac{m \Sigma_{\bar{\omega}}(m)}{F_{PS}^2 M_{PS}^2} = 1
\end{gather}
relates the PS mass and decay constant to the fermion mass and chiral condensate close to the chiral limit. It is worth noticing that all the quantities appearing in the formula above are well-defined: in particular the chiral condensate $\Sigma_{\bar{\omega}}(m)$ has been defined in eq.~\eqref{eq:sigma_sharp} with a renormalized sharp cutoff in the eigenvalue space. The GMOR relation comes from matching the IR divergences arising in the chiral limit in the two sides of the chiral Ward identity for the PS susceptibility~\eqref{eq:ward}. I studied a possible generalization of the GMOR relation to the case of theories in the conformal window. Several lessons can be learned from the computation.

It is possible to identify two ranges for the anomalous dimension $\gamma_*$ of the chiral condensate at the IR fixed point, for which the considered gauge theories behaves quite differently: large anomalous dimension $1<\gamma_*<2$ and small anomalous dimensions $0<\gamma_*<1$.

For a large anomalous dimension the PS susceptibility is IR-divergent in the chiral limit, and the (UV-regulated) chiral condensate scales like:
\begin{gather}
\Sigma_{\bar{\omega}} \simeq A m^{\frac{3-\gamma_*}{1+\gamma_*}} \ .
\end{gather}
One has to be very careful in taking the chiral limit of diverging integrals, because the chiral expansion and the integral do not commute. A clear example is eq.~\eqref{eq:chiral_Sigma}: the leading term of the chiral condensate in the chiral limit does not depend only on the Dirac eigenvalue density in the chiral limit, but keeps track of the details of the mass-deformed theory. The correct leading behavior of the PS susceptibility and chiral condensate can be extracted only after studying the effects of asymptotic scale invariance on the PS correlator and the Dirac eigenvalue density. This is where the RG equations come into play: IR scale invariance is modeled as an IR fixed point for the RG flow, close to which the theory has a certain number of regularity properties (\textit{hyperscaling hypothesis}). One learns that the pole dominance in the PS susceptibility, which is valid for spontaneously broken chiral symmetry, does not hold in the conformal window. A GMOR-like relation is obtained as for chiral symmetry breaking by matching the IR divergences in the two sides of the chiral Ward identity, but it is modified by the contribution of the continuum part of the spectrum of the PS channel in the mass-deformed theory:
\begin{gather}
\lim_{m \to 0} \frac{m \Sigma_{\bar{\omega}}(m)}{F_{PS}^2 M_{PS}^2} = 1 +
\frac{4}{\hat{f}_{PS}^2 \hat{m}_{PS}^2} \int_0^{\hat{s}_{PS}^{-1/2}} x^{-3+2\gamma_*} \hat{R}_{PS} (x,0) dx \ .
\end{gather}

For small anomalous dimension the PS susceptibility is finite in the chiral limit, and the (UV-regulated) chiral condensate scales like:
\begin{gather}
\Sigma_{\bar{\omega}} \simeq \alpha_{\bar{\omega}} m + A m^{\frac{3-\gamma_*}{1+\gamma_*}} \ .
\end{gather}
Considering the same ratio as in the GMOR relation, one gets:
\begin{gather}
\lim_{m \to 0} \frac{m \Sigma_{\bar{\omega}}(m)}{F_{PS}^2 M_{PS}^2} = +\infty \ .
\end{gather}
However this result is completely driven by the cut-off dependent part of the chiral condensate, since the term that scales with the anomalous dimension is subleading. A GMOR-like relation can be obtained by explicitly subtracting the linear term from the chiral condensate. This is done by considering derivatives with respect to the mass of the chiral Ward identity, which involve 3- and 4-point functions. Asymptotic scale invariance is not sufficient for determining the IR divergences of the derivatives of the chiral Ward identity. One has to assume that the theory that sits on the IR-fixed point is not only scale invariant, but also conformal invariant (i.e. invariant also under special conformal transformations). It is worth to notice that the existence of an IR fixed point is related to scale invariance, but does not generally imply conformal invariance. However under this extra hypothesis one gets a GMOR-like relation for the case of small anomalous dimension:
\begin{gather}
\lim_{m \to 0} \frac{m^2}{F_{PS}^2 M_{PS}^2} \left[ \frac{\Sigma_{\bar{\omega}}}{m} \right]_{sub} =1 + \frac{4}{\hat{f}_{PS}^2 \hat{m}_{PS}^2} \int_0^{\hat{s}_{PS}^{-1/2}} x^{-3+2\gamma_*} [\hat{R}_{PS} (x,0) - \hat{R}_{PS}(0,0)] dx \ ,
\end{gather}
where the \textit{sub} subscript indicates that the linear term in the mass has to be subtracted by the chiral condensate.

What are the possible applications of the found relations in lattice simulations? I have shown in sec.~\ref{sec:lattice} how the GMOR ratio $m \Sigma_{\bar{\omega}} /(F_{PS}^2 M_{PS}^2)$ can be properly defined on the lattice with Wilson fermions. This quantity is interesting because it has to be 1 in the chiral limit if chiral symmetry is spontaneously broken, and strictly larger than 1 in the conformal window (in particular it has to be infinite for small anomalous dimension). Understanding whether the computation of the GMOR ratio on the lattice is also \textit{practically} viable goes beyond the goals of this paper, and will be the subject of future investigations.

\begin{acknowledgments}
The author would like to thank Luigi Del Debbio, Roman Zwicky, Biagio Lucini and Antonio Rago for useful discussions and comments on the manuscript.
\end{acknowledgments}

\appendix

\section{Regulated chiral Ward identity}
\label{app:regulated_ward}

The goal of this appendix is to derive the regulated Ward identity in eq.~\eqref{eq:reg_ward} with the definitions in eqs.~\eqref{eq:reg_pscorr} and~\eqref{eq:reg_cc}. The starting point is the PS correlator in Euclidean spacetime. Since the fermion fields enter quadratically in the action, they can be explicitly integrated out in the functional integral:
\begin{flalign}
C_{PS}(x,m) = &
- \langle \bar{d}\gamma_5 u(x) \bar{u} \gamma_5 d(0) \rangle = \nonumber \\
= &
\langle \tr \left[ ( \Dslash + m )^{-1}(x,0) \gamma_5 ( \Dslash + m )^{-1}(0,x) \gamma_5 \right] \rangle = \nonumber \\
= &
\langle \tr \left[ ( \Dslash + m )^{-1}(x,0) ( - \Dslash + m )^{-1}(0,x) \right] \rangle
\ .
\end{flalign}
In the last line, I have used the property $\gamma_5 \Dslash \gamma_5 = -\Dslash$.

In a finite box with volume $L^4$, the massless Dirac operator at fixed gauge background has a discrete spectrum. In Euclidean spacetime it is antihermitean, hence it can be diagonalized and its eigenvalues are purely imaginary. Let $X_a(x)$ be the generic eigenvector (spin and color indices are not explicitly written), corresponding to the eigenvalue $i \omega_a$:
\begin{gather}
\Dslash X_a = i \omega_a X_a \ .
\end{gather}
The eigenvectors can be chosen to be orthogonal and normalized:
\begin{gather}
\int X_a^\dag(x) X_b(x) d^4x = \delta_{ab} \ .
\end{gather}
Both $X_a$ and $\omega_a$ depend on the gauge background. The inverse of the massive Dirac operator is:
\begin{gather}
( \pm \Dslash + m )^{-1}(x,y) = \sum_a \frac{X_a(x) X_a^\dag(y)}{\pm i \omega_a + m} \ .
\label{eq:app:invd}
\end{gather}
Plugging this formula in the PS correlator, one gets:
\begin{gather}
C_{PS}(x,m) = \lim_{L \to \infty} 
\sum_{a,b} \langle
\frac{X^\dag_a(0) X_b(0) X_b^\dag(x) X_a(x)}{(i \omega_a + m)(-i \omega_b + m)}
\rangle \ .
\end{gather}

The regulated PS correlator is obtained by introducing a regulator $\R_{\omega_\infty}(\omega)$ which cuts the eigenvalues above $\omega_\infty$. One can choose for instance $\R_{\omega_\infty}(\omega) = \exp ( - \omega^2/\omega_\infty^2 )$. When the cutoff is removed, the full PS correlator is recovered.
\begin{gather}
C_{PS}^{\R}(x,m,\omega_\infty) = \lim_{L \to \infty} \sum_{a,b} \langle \frac{X_a^\dag(0) X_b(0) X_b^\dag(x) X_a(x)}{(i \omega_a + m)(-i \omega_b + m)} \R_{\omega_\infty} (\omega_a) \R_{\omega_\infty} (\omega_b) \rangle \ .
\\
\lim_{\omega_\infty \to \infty} C_{PS}^{\R}(x,m,\omega_\infty) = C_{PS}(x,m) \ .
\end{gather}

The regulated PS susceptibility is obtained by integrating the regulated PS correlator over spacetime. One can use the orthogonality of eigenvectors and the translation invariance to get:
\begin{flalign}
& \int d^4x \ C_{PS}^{\R}(x,m,\omega_\infty) = \nonumber \\
& = \lim_{L \to \infty} \frac{1}{L^4} \int d^4xd^4y \ 
\sum_{a,b} \langle \frac{X_a^\dag(y) X_b(y) X_b^\dag(x) X_a(x)}{(i \omega_a + m)(-i \omega_b + m)} \R_{\omega_\infty} (\omega_a) \R_{\omega_\infty} (\omega_b) \rangle = \nonumber \\
& =
\lim_{L \to \infty} \frac{1}{L^4}
\sum_{a} \langle \frac{\R_{\omega_\infty}^2(\omega_a)}{\omega_a^2 + m^2} \rangle
\ .
\end{flalign}
The last line of the previous equation is related to the chiral condensate. In fact integrating out the fermion fields explicitly, using the representation~\eqref{eq:app:invd} and the reality of the chiral condensate, one can show that the chiral condensate is:
\begin{flalign}
\lim_{L \to \infty} \frac{1}{L^4} \int d^4x \langle \bar{u}u(x) \rangle = &
- \lim_{L \to \infty} \frac{1}{L^4} \int d^4x \langle \tr ( \Dslash + m )^{-1}(x,x) \rangle = \nonumber \\
%= &
%- \lim_{L \to \infty} \frac{1}{L^4} \sum_a \int d^4x \langle \frac{X_a(x) X_a^\dag(x)}{i \omega_a + m} \rangle = \\
%= &
%- \lim_{L \to \infty} \frac{1}{L^4} \sum_a \langle \frac{1}{i \omega_a + m} \rangle = \\
%= &
%- \lim_{L \to \infty} \frac{1}{L^4} \sum_a \langle \frac{-i \omega_a + m}{\omega_a^2 + m^2} \rangle = \\
= &
- \lim_{L \to \infty} \frac{m}{L^4} \sum_a \langle \frac{1}{\omega_a^2 + m^2} \rangle \ .
\end{flalign}
It is well known that this expression is UV-divergent. One can define a regulated (absolute value of the) chiral condensate as:
\begin{gather}
\Sigma_{\R} = \lim_{L \to \infty} \frac{m}{L^4} \sum_a \langle \frac{\R_{\omega_\infty}^2 (\omega_a)}{\omega_a^2 + m^2} \rangle \ .
\end{gather}
Putting this together with the expression for the regulated PS susceptibility, one finally gets the regulated Ward identity:
\begin{gather}
\int d^4x \ C_{PS}^{\R}(x,m,\omega_\infty) = \frac{\Sigma_{\R}}{m} \ .
\end{gather}

\section{RG equations}
\label{app:RG}

\subsection{General formulae}

In this appendix I want to derive explicitly eq.~\eqref{eq:RG_O_5} from the RG equation:
\begin{gather}
\left\{
\mu_0 \frac{\partial}{\partial \mu_0} + \beta(g) \frac{\partial}{\partial g} - \gamma(g) m \frac{\partial}{\partial m} - \gamma_O(g)
\right\} O(E,g,m,\mu_0) = 0 \ .
\label{eq:app:RG_O_1}
\end{gather}

The first step for solving the RG equation is to introduce the running coupling $\bar{g}$ and mass $\bar{m}$, and the renormalization factor $\Z_O$, that satisfy the following equations (\textit{RG flow}):
\begin{gather}
\mu \frac{d}{d\mu}\bar{g}(\mu) = \beta(\bar{g}(\mu)) \ ,
\label{eq:app:gbar}
\\
\mu \frac{d}{d\mu}\bar{m}(\mu) = -\gamma(\bar{g}(\mu)) \bar{m}(\mu) \ ,
\label{eq:app:mbar}
\\
\mu \frac{d}{d\mu} \log \Z_O(\mu) = -\gamma_O(\bar{g}(\mu)) \ ,
\label{eq:app:zo}
\\
\bar{g}(\mu_0) = g \ , \qquad \bar{m}(\mu_0) = m \ , \qquad \Z_O(\mu_0)=1 \ .
\label{eq:app:initial}
\end{gather}

Using the derivative chain-rule, one can easily show that the RG equation is equivalent to the following one:
\begin{gather}
\frac{d}{d\mu} \left[ \Z_O(\mu) O(E,\bar{g}(\mu),\bar{m}(\mu),\mu) \right] = 0 \ .
\label{eq:app:RG_O_2}
\end{gather}
This equation is readily integrated in terms of the initial conditions of the RG flow:
\begin{gather}
\Z_O(\mu) O(E,\bar{g}(\mu),\bar{m}(\mu),\mu) = O(E,g,m,\mu_0) \ .
\label{eq:app:RG_O_3}
\end{gather}

Before proceeding further, one needs to go back to equations of the RG flow. Eq.~\eqref{eq:app:gbar} for the running coupling can be solved implicitly. One can use asymptotic freedom and the perturbative expansion for the $\beta$ function:
\begin{gather}
\beta(g) = -\beta_0 g^3 + \beta_1 g^5 + O(g^7) \ ,
\end{gather}
in order to write the solutions of eq.~\eqref{eq:app:gbar} as function of a single integration constant $\Lambda$ (which depends on $\mu_0$ and $g$):
\begin{gather}
\frac{\Lambda}{\mu} = e^{-\frac{1}{2\beta_0 \bar{g}^2(\mu)}} \bar{g}(\mu)^{\frac{\beta_1}{\beta_0^2}} \exp \left\{ - \int_0^{\bar{g}(\mu)} \left( \frac{1}{\beta(g)} + \frac{1}{\beta_0 g^3} + \frac{\beta_1}{\beta_0^2 g} \right) dg \right\} \ .
\label{eq:app:run_g}
\end{gather}
The running coupling depends on the scale $\mu$ but also on the initial condition via the integration constant $\Lambda$. When needed, I will use the notation $\bar{g}(\mu;\Lambda)$. Eq.~\eqref{eq:app:mbar} for the running mass can be integrated explicitly. However also in this case the initial condition can be traded for an integration constant $M$ (RG-invariant mass) that satisfies $\bar{m}(M)=M$. The running mass can be written as:
\begin{gather}
\bar{m}(\mu;\Lambda,M) = \exp \left\{ -\int_{\bar{g}(M;\Lambda)}^{\bar{g}(\mu;\Lambda)} \frac{\gamma(z)}{\beta(z)} dz \right\} M \ .
\label{eq:app:run_m}
\end{gather}
Notice once again that the running coupling and mass do not depend explicitly on $\mu_0$ once they are expressed in terms of the integration constants $\Lambda$ and $M$. Also eq.~\eqref{eq:app:zo} for the renormalization constant can be integrated explicitly:
\begin{gather}
\Z_O(\mu;\mu_0,\Lambda) = \exp \left\{ -\int_{\bar{g}(\mu_0;\Lambda)}^{\bar{g}(\mu;\Lambda)} \frac{\gamma_O(z)}{\beta(z)} dz \right\} \ .
\label{eq:app:run_zo}
\end{gather}
However a residual dependence on the renormalization scale $\mu_0$ survives in this case.

One can go back to eq.~\eqref{eq:app:RG_O_3}, choose $\mu=E$, and write explicitly the dependence on $\mu_0$, $\Lambda$, $M$:
\begin{gather}
O(E,g,m,\mu_0) = \Z_O(E;\mu_0,\Lambda) O(E,\bar{g}(E;\Lambda),\bar{m}(E;\Lambda,M),E) \ .
\label{eq:app:RG_O_4}
\end{gather}
Being $d_O$ the engineered dimension in mass of the observable $O$, and being $\Z_O$ dimensionless, from dimensional analysis one gets:
\begin{gather}
O(E,\bar{g}(E;\Lambda),\bar{m}(E;\Lambda,M),E) = E^{d_O} \tilde{O}\left( \frac{M}{E} , \frac{E}{\Lambda} \right) \ , \\
\Z_O(E;\mu_0,\Lambda) = \tilde{\Z}_O \left( \frac{\mu_0}{\Lambda} , \frac{E}{\Lambda} \right) \ .
\label{eq:app:tildezo}
\end{gather}
Plugging these last two equations into eq.~\eqref{eq:app:RG_O_4}, one finally gets:
\begin{gather}
O(E,g,m,\mu_0) =
\tilde{\Z}_O \left( \frac{\mu_0}{\Lambda} , \frac{E}{\Lambda} \right)
E^{d_O} \tilde{O}\left( \frac{M}{E} , \frac{E}{\Lambda} \right)
\ .
\label{eq:app:RG_O_5}
\end{gather}

\subsection{Mass-deformed IR-conformal theory}

IR scale invariance is recovered in the chiral limit if the beta function has a zero in $g=g_*$ and the coupling constant runs into it in the IR limit $\bar{g}(\mu \to 0) = g_*$. I will assume regularity of the beta function and the anomalous dimensions around $g=g_*$ (hyperscaling hypothesis):
\begin{gather}
\beta(g) = \beta_* (g-g_*) + O((g-g_*)^2) \ , \\
\gamma(g) = \gamma_* + O(g-g_*) \ , \\
\gamma_O(g) = \gamma_*^{(O)} + O(g-g_*) \ .
\end{gather}
The goal of this subsection is to specialize the formulae of the previous one to this particular case.

The running coupling can be expanded at small renormalization scale $\mu \ll \Lambda$:
\begin{gather}
\bar{g}(\mu) \simeq g_* - A \left( \frac{\mu}{\Lambda} \right)^{\beta_*}  \ .
\end{gather}
Notice that $\beta_* >0$ since $g_*$ is an IR fixed point, and $A>0$ since we are considering asymptotically-free solutions with $\bar{g}(\infty)=0$.

It is useful to introduce the following function:
\begin{gather}
Z_O(\mu/\Lambda)
= \left( \frac{\mu}{\Lambda} \right)^{-\gamma_*^{(O)}} \left[ \frac{g_*-\bar{g}(\mu/\Lambda)}{A} \right]^{\frac{\gamma_*^{(O)}}{\beta_*}}
\exp \left\{ - \int_{\bar{g}(\mu/\Lambda)}^{g_*} \left( \frac{\gamma_O(z)}{\beta(z)} + \frac{\gamma_*^{(O)}}{\beta_*(g_*-z)} \right) dz \right\} \ ,
\label{eq:app:zo_IRCT}
\end{gather}
which is normalized in such a way that $Z_O(0)=1$, and in terms of which the renormalization factor is written as:
\begin{gather}
\tilde{\Z}_O \left(\frac{\mu_0}{\Lambda}, \frac{\mu}{\Lambda} \right) = \frac{Z_O(\mu_0/\Lambda)}{Z_O(\mu/\Lambda)} \left( \frac{\mu_0}{\mu} \right)^{\gamma_*^{(O)}}\ .
\label{eq:app:tildezo_IRCT}
\end{gather}

One can plug the $Z_O$ function into eq.~\eqref{eq:app:RG_O_5}, and define $\hat{O}=\tilde{O}/Z_O(E/\Lambda)$ in order to get:
\begin{gather}
O(E,g,m,\mu_0) =
Z_O \left( \frac{\mu_0}{\Lambda} \right) \mu_0^{\gamma_*^{(O)}}
E^{d_O-\gamma_*^{(O)}} \hat{O}\left( \frac{M}{E} , \frac{E}{\Lambda} \right)
\ .
\label{eq:app:RG_O_6}
\end{gather}

Replacing $\gamma_O$ with $\gamma$ in eqs.~\eqref{eq:app:run_zo}, \eqref{eq:app:tildezo}, \eqref{eq:app:zo_IRCT} and \eqref{eq:app:tildezo_IRCT}, one can define the quantities $\Z$, $\tilde{\Z}$, $Z$ in analogy with $\Z_O$, $\tilde{\Z}_O$, $Z_O$. The running mass defined in eq.~\eqref{eq:app:run_m} can be written as:
\begin{gather}
\bar{m}(\mu) = \tilde{\Z}^{-1} \left( \frac{\mu}{\Lambda} , \frac{M}{\Lambda} \right) M
= \frac{Z(M/\Lambda)}{Z(\mu/\Lambda)} \mu^{-\gamma_*} M^{1+\gamma_*} \ .
\end{gather}
which displays the typical power law of IR-conformal theories. Notice that $\Z$ is the renormalization factor of the chiral condensate (therefore the mass renormalizes with $\Z^{-1}$).

\bibliography{gmor}

\end{document}